\documentclass[12pt,a4paper]{article}
\usepackage{amsfonts,latexsym}
\usepackage{amsmath,amssymb}
\usepackage{graphicx,color}
\usepackage{multirow}
\usepackage{slashbox}
\newcommand{\nn}{\nonumber}

\begin{document}
\vspace{12mm}

\begin{center}
{{{\Large {\bf Black holes in Gauss-Bonnet and Chern-Simons-scalar theory}}}}\\[10mm]

{Yun Soo Myung$^a$\footnote{e-mail address: ysmyung@inje.ac.kr} and De-Cheng Zou$^{a,b}$\footnote{e-mail address: dczou@yzu.edu.cn}}\\[8mm]

{${}^a$Institute of Basic Sciences and Department  of Computer Simulation, \\ Inje University, Gimhae 50834, Korea\\[0pt] }

{${}^b$Center for Gravitation and Cosmology and College of Physical Science and Technology, Yangzhou University, Yangzhou 225009, China\\[0pt]}
\end{center}
\vspace{2mm}

\begin{abstract}
We carry out  the stability analysis of the Schwarzschild black hole
in Gauss-Bonnet and  Chern-Simons-scalar theory. Here, we
introduce two quadratic scalar couplings ($\phi_1^2,\phi_2^2$) to Gauss-Bonnet and  Chern-Simons terms, where
the former term is parity-even, while the latter one is parity-odd.
The perturbation equation for the scalar
$\phi_1$ is the Klein-Gordon equation with an effective mass, while the perturbation equation for
$\phi_2$ is coupled to the parity-odd metric perturbation,
providing a system of  two coupled  equations.  It turns
out that the Schwarzschild black hole is unstable against  $\phi_1$
perturbation, leading to  scalarized black holes, while the  black hole is stable against  $\phi_2$ and metric perturbations, implying no scalarized black holes.

\end{abstract}

\newpage
\renewcommand{\thefootnote}{\arabic{footnote}}
\setcounter{footnote}{0}


\section{Introduction}
Black holes with scalar hair (spontaneous scalarization) are obtained  dynamically in a different class of models where
the scalar field   couples either to the topological invariant, Gauss-Bonnet term~\cite{Antoniou:2017acq,Doneva:2017bvd,Silva:2017uqg} or to Maxwell kinetic term~\cite{Stefanov:2007eq,Doneva:2010ke,Herdeiro:2018wub} with a coupling function $f(\phi)$.
In this case, the appearance of static scalarized solutions is closely connected to the instability of  Schwarzschild~\cite{Myung:2018iyq} or Reissner-Nordstr\"{o}m  solution~\cite{Myung:2018vug} without scalar hair. Here, the coupling constant $\alpha$ plays the role of a spectral parameter (effective mass) in the linearized equation.
However, the other topological invariant of  Chern-Simon term does not activate the static scalarized black hole, but it could develop the
scalarized Kerr solution with linear coupling~\cite{Delsate:2018ome} and scalarized Schwarzschild-NUT solution with quadratic coupling~\cite{Brihaye:2018bgc}. It was known that the Chern-Simons term is a parity-odd (violating) one and  thus,  non-rotating black holes are not modified because
these are parity-even solution~\cite{Ayzenberg:2013wua}. The Kerr and  Schwarzschild-NUT solutions are parity-odd and thus, they do acquire modifications from the Chern-Simons term.
We note that Schwarzschild black hole was stable in the Einstein-Chern-Simons gravity with linear coupling~\cite{Molina:2010fb,Moon:2011fw,Kimura:2018nxk,Macedo:2018txb}, which  may explain why scalarized static  black holes do not exist.

Similarly, cosmology with the Gauss-Bonnet term could affect the background evolution of parity-even and results in the tensor-to-scalar ratio $r$, leading to violation of the consistency relation~\cite{Satoh:2008ck,Satoh:2010ep}.  On the other hand, the Chern-Simons term did not appear in the background and scalar perturbations, but this appears in tensor perturbations as circularly polarized modes because these modes belong to the parity-odd case.

An effective action including  both topological invariants with different linear couplings could be obtained from some superstring models~\cite{Antoniadis:1993jc} and the heterotic strings~\cite{Cano:2019ore}.
Inspired by this, we introduce a new action (\ref{Action}) which includes  both topological invariants with different quadratic couplings.
We will use this new action to investigate the stability analysis of the Schwarzschild black hole  which is closely related to the spontaneous scalarization of black hole.
This will reveal the hidden roles of two topological invariants on the scalarization process.

To make all things clear, we mention our notations. We will use natural units of $G=c=\hbar=1$ with  signature $(-,+,+,+)$. The Riemann, Ricci tensor,
and Levi-Civita tensor are defined by
\begin{eqnarray}
R^{\rho}_{~\sigma\mu\nu}=\partial_{\mu}\Gamma^{\rho}_{\nu\sigma}
-\partial_{\nu}\Gamma^{\rho}_{\mu\sigma}+
\Gamma^{\rho}_{\mu\lambda}\Gamma^{\lambda}_{\nu\sigma}-
\Gamma^{\rho}_{\nu\lambda}\Gamma^{\lambda}_{\mu\sigma},~
R_{\mu\nu}=R^{\rho}_{~\mu\rho\nu},~
\epsilon^{tr\varphi_1\varphi_2}=\frac{1}{\sqrt{-g}}. \nonumber
\end{eqnarray}

\section{GBCSS theory}
We  introduce newly the Gauss-Bonnet and Chern-Simons-scalar (GBCSS) theory in four dimensions whose action is given by
\begin{eqnarray}
S_{\rm GBCSS}=\frac{1}{16 \pi}\int d^4 x\sqrt{-g} \Bigg[
R-\frac{1}{2}(\partial \phi_1)^2-\frac{1}{2}(\partial \phi_2)^2+ \alpha\phi_1^2 R^2_{\rm GB}+\beta\phi_2^2{}^{*}RR
\Bigg]\label{Action}
\end{eqnarray}
where $R^2_{\rm GB}$ is the Gauss-Bonnet term
\begin{equation}
R^2_{\rm GB}=R^2-4R_{\mu\nu}R^{\mu\nu}+R_{\mu\nu\rho\sigma}R^{\mu\nu\rho\sigma}
\end{equation}
and ${}^{*}RR$ is  the
Chern-Simons term
\begin{equation}
{}^{*}RR={}^{*}R^{\eta~\mu\nu}_{~\xi}R^\xi_{~\eta\mu\nu}.
\end{equation}
Here the dual Ricci tensor is defined by
\begin{eqnarray}
{}^{*}R^{\eta~\mu\nu}_{~\xi}=\frac{1}{2}\epsilon^{\mu\nu\rho\sigma}R^{\eta}_{~\xi\rho\sigma}.
\end{eqnarray}
From now on, we call the EGBS (ECSS) theory for the case of $\beta=0(\alpha=0)$.
It is worth noting that the mass dimensions of coupling constants are given by $[\alpha]=[\beta]=-2$.
Although both $R^2_{\rm GB}$ and ${}^{*}RR$ are topological (total derivatives) in four dimensions, but
they become dynamical due to the coupling to the scalars. However, there exists a  difference between them.
The Gauss-Bonnet term affects the property of the static solutions of parity-even source, while the Chern-Simons term
gives different results only in the rotating solution of parity-odd source. This explains that the scalarized Schwarzschild black holes could be found from
the Gauss-Bonnet coupling,  but they might  not be  found  from the Chern-Simons coupling. In other words,  the Chern-Simons term
does not activate a scalar monopole field in the static black hole spacetime.  Also, the appearance of scalarized black holes is closely related to
the instability of scalar-free black hole. Hence, it is important to investigate the stability of black holes without scalar hair in the GBCSS theory.

Varying for $g_{\mu\nu},~\phi_1$, and $\phi_2$ lead to the
three equations as
\begin{eqnarray}
&&G_{\mu\nu}=\frac{1}{2}\partial_{\mu}\phi_1\partial_{\nu}\phi_1+\frac{1}{2}\partial_{\mu}\phi_2\partial_{\nu}\phi_2
- \frac{1}{4}g_{\mu\nu}[(\partial \phi_1)^2+(\partial \phi_2)^2] \nonumber \\
&&\hspace*{3em}-4\alpha \nabla^\rho \nabla^\sigma (\phi_1^2)P_{\mu\rho\nu\sigma} -4\beta C_{\mu\nu},\label{eq-1}\\
&&\nabla^2\phi_1=-2\alpha R^2_{\rm GB} \phi_1,\label{eq-2}\\
&&\nabla^2\phi_2=-2\beta~ {}^{*}RR \phi_2 , \label{eq-3}
\end{eqnarray}
where $P_{\mu\rho\nu\sigma}$-tensor takes the form
\begin{equation}
P_{\mu\rho\nu\sigma}=R_{\mu\rho\nu\sigma}+g_{\mu\sigma}R_{\nu\rho}-g_{\mu\nu}R_{\rho\sigma}+g_{\nu\rho}R_{\mu\sigma}-g_{\rho\sigma}R_{\mu\nu}
+\frac{R}{2}(g_{\mu\nu}g_{\rho\sigma}-g_{\mu\sigma}g_{\nu\rho}),
\end{equation}
which corresponds to  the divergence-free part
of the Riemann tensor ($\nabla_\mu P^{\mu}_{~\rho\nu\sigma}=0$).
Here the Cotton tensor  $C_{\mu\nu}$ is given by
\begin{eqnarray}\label{cotton}
C_{\mu\nu}=\nabla_{\rho}(\phi_2^2)~\epsilon^{\rho\sigma
\gamma}_{~~~~(\mu}\nabla_{\gamma}R_{\nu)\sigma}+\frac{1}{2}\nabla_{\rho}\nabla_{\sigma}
(\phi_2^2)~\epsilon_{(\nu}^{~~\rho \gamma
\delta}R^{\sigma}_{~~\mu)\gamma \delta}.
\end{eqnarray}
Using  the trace  of (\ref{eq-1})
\begin{equation}
R=\frac{1}{2} (\partial \phi_1)^2+\frac{1}{2} (\partial \phi_2)^2-4\alpha\nabla^\rho\nabla^\sigma (\phi_1^2) G_{\rho\sigma},
\end{equation}
 we  rewrite (\ref{eq-1}) as  the Ricci-tensor equation
\begin{eqnarray}\label{eq-4}
&&R_{\mu\nu}=\frac{1}{2}\partial_{\mu}\phi_1\partial_{\nu}\phi_1+\frac{1}{2}\partial_{\mu}\phi_2\partial_{\nu}\phi_2 \nonumber \\
&&\hspace*{3em}-4\alpha \nabla^\rho\nabla^\sigma (\phi_1^2)\Big[R_{\mu\rho\nu\sigma}+g_{\mu\sigma}R_{\nu\rho}+g_{\nu\rho}R_{\mu\sigma}\Big]-4\beta C_{\mu\nu} \label{eq-5}
\end{eqnarray}
which may be  suitable for analyzing the stability of the black holes.

Choosing  the background quantities (by denoting  the ``overbar'')
 \begin{equation}
\bar{\phi}_1=0,~\bar{\phi}_2={\rm const},
\end{equation}
Eqs.(\ref{eq-1})-(\ref{eq-3}) admit the spherically symmetric Schwarzschild spacetime
\begin{eqnarray}
ds_{\rm S}^2&=&\bar{g}_{\mu\nu}dx^{\mu}dx^{\nu}\nn\\
&=&-f(r)dt^2+\frac{dr^2}{f(r)}+r^2(d\varphi_1^2+\sin^2\varphi_1
d\varphi_2^2) \label{sch-sol}
\end{eqnarray}
with the metric function
\begin{equation} f(r)=1-\frac{2M}{r}=1-\frac{r_+}{r}.
 \end{equation}
In this case, one has
\begin{equation}
\bar{R}^2_{\rm GB}=\frac{48 M^2}{r^6},~{}^{*}\bar{R}\bar{R}=0, ~\bar{C}_{\mu\nu}=0,~\bar{R}=\bar{R}_{\mu\nu}=0,~ \bar{R}_{\mu\rho\nu\sigma}\not=0.
\end{equation}

Now let us  introduce the perturbations
around the background  as
\begin{eqnarray} \label{m-p}
g_{\mu\nu}=\bar{g}_{\mu\nu}+h_{\mu\nu},~\phi_1=0+\delta\phi_1,~~~\phi_2=\bar{\phi}_2+\delta\phi_2.
\end{eqnarray}
The linearized equation to (\ref{eq-4}) can be written by
\begin{eqnarray}\label{pertg}
\delta R_{\mu\nu}(h)=-4\beta \delta C_{\mu\nu}(\delta \phi_2)
\end{eqnarray}
where  $\delta R_{\mu\nu}(h)$ and $\delta C_{\mu\nu}(\delta \phi_2)$ take the forms
\begin{eqnarray}\label{cottonp0}
\delta
R_{\mu\nu}(h)&=&\frac{1}{2}\left(\bar{\nabla}^{\gamma}\bar{\nabla}_{\mu}
h_{\nu\gamma}+\bar{\nabla}^{\gamma}\bar{\nabla}_{\nu}
h_{\mu\gamma}-\bar{\nabla}^2h_{\mu\nu}-\bar{\nabla}_{\mu} \bar{\nabla}_{\nu} h\right),\label{cottonp1}\\
\label{cottonp2}\delta
C_{\mu\nu}(\delta \phi_2)&=&\bar{\phi}_2\bar{\nabla}_{\rho}\bar{\nabla}_{\sigma}
~\delta\phi_2~\epsilon_{(\nu}^{~~\rho \gamma
\delta}\bar{R}^{\sigma}_{~~\mu)\gamma \delta}.
\end{eqnarray}
We observe from (\ref{eq-3}) and (\ref{cottonp2})  that $\bar{\phi}_1=\bar{\phi}_2=0$ gives the same solution (\ref{sch-sol}), but it gives no coupling to the perturbed Einstein equation. Also, we note from (\ref{eq-2}) that `$\bar{R}^2_{\rm GB}\not=0$' does not admit $\bar{\phi}_1={\rm const}$ solution for the $\phi_1^2$ coupling.
From Eqs.(\ref{eq-2}) and (\ref{eq-3}), we obtain
the linearized scalar  equation for $\delta \phi_1$
\begin{eqnarray}\label{pertphi-1}
\Big(\bar{\nabla}^2+\frac{96 \alpha M^2}{r^6}\Big)\delta\phi_1=0,
\end{eqnarray}
while $\delta \phi_2$ couples to the  metric perturbation $h_{\mu\nu}$ as
\begin{eqnarray}\label{pertphi-2}
\bar{\nabla}^2\delta\phi_2+4\beta \bar{\phi}_2
\epsilon^{\mu\nu\rho\sigma}\bar{R}^{\eta}_{~\xi\mu\nu}
\bar{\nabla}_{\rho}\bar{\nabla}_{\eta}h^{\xi}_{\sigma}=0.
\end{eqnarray}
Before we proceed, we would like to focus on  two linearized theories.
In the single scalar coupling of $\phi_1=\phi_2=\phi$, choosing $\bar{\phi}=0$ provides the Schwarzschild solution.
However, the linearized equations around the black hole lead to
\begin{eqnarray}
 && \delta R_{\mu\nu} =0, \label{phie-eq1} \\
 &&  \Big(\bar{\nabla}^2+\frac{48 \alpha M^2}{r^6}\Big)\delta\phi=0,\label{phie-eq2}
\end{eqnarray}
which lead  to the linearized version for the EGBS theory
because the linearized Cotton term ($\delta C_{\mu\nu}$) decouples from (\ref{phie-eq1}).

For the other case of linear couplings ($\alpha \phi_1$ and $\beta \phi_2$) with possessing a shift symmetry of $\phi_1 \to \phi_1+c_1$ and $\phi_1 \to \phi_2+c_2$, the Schwarzschild black hole (\ref{sch-sol}) is not a solution because of non-zero Gauss-Bonnet term.
In the ECSS (linear coupling) theory,
it turned out that the Schwarzschild black hole is stable against the metric and scalar perturbations~\cite{Molina:2010fb,Moon:2011fw,Kimura:2018nxk,Macedo:2018txb}.

\section{Stability Analysis}
\subsection{Instability of $\delta \phi_1$}
We wish to start with stability analysis by noting that $\delta \phi_1$ is completely decoupled from other fields.
Using the tortoise coordinate ($r^{*}=\int dr/f(r)$) and $\delta\phi_1=\frac{u(r)}{r}Ye^{-i\omega t}$ with $Y\equiv Y_{lm}(\varphi_1,\varphi_2)$ spherical harmonics,
the linearized  equation (\ref{pertphi-1})
becomes
\begin{eqnarray}
\frac{d^2u}{dr^{*2}}+\Big[\omega^2-V_{\rm u}(r)\Big]u(r)&=&0,
\end{eqnarray}
where the  potential $V_{\rm u}(r)$ is given by
\begin{eqnarray} \label{phi1-pot}
V_{\rm u}(r)=f(r)\Big(\frac{\lambda}{r^2}+\frac{2M}{r^3} -\frac{96 \alpha M^2}{r^6}\Big)
\end{eqnarray}
with $\lambda=l(l+1)$. We  note that the last term in (\ref{phi1-pot}) plays the role of an effective mass with $[\alpha^{-1}]=2$.
This term contributes to potential large negatively near the horizon, while its contribution becomes small neglectfully as $r$ increases.
From now on, we focus on the $l=0$-mode of $u(r)$ because it is responsible for analyzing the stability analysis and obtaining scalarized black holes.
The $s(l=0)$-mode potential $V_{\rm u}^{l=0}(r)$ develops negative region outside the  horizon, depending the value of  coupling constant $\alpha$.
The sufficient condition for the instability is given by
\begin{equation}
\int^\infty_{r_+=2M} \Big[\frac{V_{\rm u}^{l=0}(r)}{f(r)} \Big] dr<0
\to \frac{M^2}{\alpha}<\frac{12}{5} \to 0<\frac{r_+}{\sqrt{\alpha}}<3.098. \label{sc-ins}
\end{equation}
However,  (\ref{sc-ins}) is not a necessary and sufficient condition for instability.
To determine the threshold of instability precisely, one has to solve the linearized equation numerically
\begin{eqnarray} \label{phi1-seq1}
\frac{d^2u}{dr^{*2}}-\Big[\Omega^2+V_{\rm u}^{l=0}(r)\Big]u(r)&=&0,
\end{eqnarray}
which may allow an exponentially growing mode of $e^{\Omega t}$ as an unstable mode.
We obtain the unstable bound for the scalar perturbation
\begin{equation}
 0<\frac{r_+}{\sqrt{\alpha}}<3.321, \label{sc-thr}
\end{equation}
which implies that the threshold of instability is determined  by  $1/\sqrt{\alpha}=1/\sqrt{\alpha_{\rm th}}=3.321$, being greater than 3.098 (sufficient condition for instability).
That is, the unstable bound for the coupling constant $\alpha$ is given by
\begin{equation}
\alpha> \frac{r_+^2}{11.03},
\end{equation}
which is smaller than the sufficient condition for instability ($\alpha>r_+^2/9.6$).

On the other hand, we solve  the linearized equation (\ref{phi1-seq1}) with $\Omega=0$ to  find a discrete spectrum of the coupling constant $\alpha$ when  obtaining static solutions:
$1/\sqrt{\alpha_s}=r_+/\sqrt{\alpha}\in[3.321,~1.281,~0.792,~0.571,~\cdots]$ where we identify the first value with the threshold of instability. These solutions are labelled by the order number $n=0$ (fundamental branch), $1,~2,~3,~\cdots$ (excited branches) which is identified with the number of nodes for $\delta \phi_1(z)=u(z)/z$ with $z=r/r_+$. Actually, it may represent the $n=0,~1,~2,~3,~4,~\cdots$ scalarized black holes found  when solving  full equations (\ref{eq-1})-(\ref{eq-3}). This implies that the appearance of $n=0$ scalarized black hole is closely related to the threshold of instability for Schwarzschild black hole.

\subsection{Stability of $\delta \phi_2$ and $h_{\mu\nu}$}
The metric perturbation $h_{\mu\nu}$ is  classified according to
the transformation properties under parity, namely odd sector
($h_0,~h_1$) and even sector ($H_0,~H_1,~H_2,K$) as
\begin{eqnarray} h_{\mu\nu}(t,r,\varphi_1,\varphi_2)= e^{-i\omega t}\left(
\begin{array}{cccc}
H_0(r)Y & H_1(r)Y &
-h_0(r)\frac{\partial_{\varphi_2}Y}{\sin\varphi_1}& h_0(r)\sin\varphi_1
\partial_{\varphi_1}Y
\cr
* & H_2(r)Y
&-h_1(r)\frac{\partial_{\varphi_2}Y}{\sin\varphi_1}& h_1(r) \sin\varphi_1
\partial_{\varphi_1}Y \cr
*&
*& r^2YK(r) & 0
\cr * &* & * &
r^2\sin^2\varphi_1YK(r)
\end{array}
\right) \label{pmetric}
\end{eqnarray}
with  $*$ symmetrizations. The
form of $\delta\phi_2$ is given by
\begin{eqnarray}\label{perttheta1}
\delta\phi_2=\frac{\psi(r)}{r}Ye^{-i\omega t}.
\end{eqnarray}
Plugging  Eqs. (\ref{pmetric}) and  (\ref{perttheta1}) into
Eq.(\ref{pertg}), we find
 ten perturbation equations as appeared in Appendix.
It is important to note that  ten perturbation equations imply twenty constraints like
\begin{eqnarray}
E_{i}=0,~~~O_{i}=0,~~~ {\rm for}~i=1,\cdots,10, \label{eo-eqs}
\end{eqnarray}
where $E_{i}$ with $i=1,\cdots,10$ are functions of
($H_0,~H_1,~H_2,~K$) and $O_{i}$ with $i=1,\cdots,10$ are
functions of ($h_0,~h_1,~\psi$).
This implies  that ten perturbation equations can be classified  into two
parties: odd-parity equations ($\{O_{i}=0\}$) and even-parity equations  ($\{E_{i}=0\}$).
We emphasize that there are  couplings between $h_{\mu\nu}$ and $\delta \phi_2$ in the odd-parity  equations.

First of all, we consider the even-parity ($\{E_{i}=0\}$)  because it corresponds to even-parity sector of Einstein gravity ($\delta R_{\mu\nu}=0$).
It is well known that  this case  reduces to a single second-order equation for  a field defined by
\begin{equation}\label{calM}
\hat{M}=\frac{1}{p(r)q(r)-h(r)}\left\{p(r)K(r)
-\frac{H_1}{\omega}\right\},
\end{equation}
where
\begin{eqnarray}
q(r)&=&\frac{\tilde{\lambda}(\tilde{\lambda}+1)r^2+3\tilde{\lambda}
Mr+6M^2}{r^2(\tilde{\lambda} r+3M)},~~ h(r)=\frac{i(-\tilde{\lambda}
r^2+3\tilde{\lambda}
Mr+3M^2)}{(r-2M)(\tilde{\lambda} r+3M)},\nn\\
~~~p(r)&=&-\frac{ir^2}{r-2M},~~ \tilde{\lambda}=\frac{\lambda}{2}-1.
\end{eqnarray}
Here, we
obtain the Zerilli equation~\cite{Zerilli:1970se}
\begin{eqnarray} \label{evenz}
\frac{d^2{\hat{M}}}{dr^{*2}}+\Big[\omega^2-V_{\rm Z}(r)\Big]{\hat{M}}=0,
\end{eqnarray}
where the Zerilli potential is given by
\begin{equation}
V_{\rm Z}(r)=f(r)
\Bigg[\frac{2{\tilde{\lambda}}^2(\tilde{\lambda}+1)r^3+6\tilde{\lambda}^2Mr^2
+18\tilde{\lambda} M^2 r+18M^3} {r^3(\tilde{\lambda} r+3M)^2}\Bigg].
\end{equation}
All potentials $V_{\rm Z}(r)$ for $l\ge 2(\tilde{\lambda}\ge2)$ are always positive for whole range of
$r_+ \le r \le \infty$, which implies that the even-parity
perturbation is stable.

On the other hand,  for odd-parity sector ($\delta R_{\mu\nu}=-4\beta \delta C_{\mu\nu}$),
the first five equations ($\{O_{i}=0\},i=1,\cdots,5$) provide three relevant equations with $\tilde{\beta}=8\bar{\phi}_2\beta$:
\begin{eqnarray}
&&  O_1=0(O_2=0)\nn\\
&&r^3(-4M+\lambda r)h_0-r
f\Big(2i \omega r^4 h_1-6\tilde{\beta} M\psi+i\omega r^5h_1^{\prime} \nn \\
&&\hspace*{3em}+3\tilde{\beta} Mr\psi^{\prime} +r^5h_0^{\prime\prime}\Big) =0,\label{feq1}\\
&&O_3=0(O_4=0)\nn\\
&&
-i\omega r^3\Big(2h_0-i\omega rh_1-rh_0^{\prime}\Big)+r^2f(\lambda-2)h_1+3 i\tilde{\beta} \omega M\psi=0,\label{feq2}\\
&&O_5=0\nn\\
&&i \omega r^3h_0-(2M-r)\Big\{2Mh_1-(2M-r)rh_1^{\prime}\Big\}=0.\label{feq3}
\end{eqnarray}
We note that all remaining equations $O_i=0$ with $i=6,\cdots,10$ are
redundant. Introducing  a new field $Q=2fh_1/(\tilde{\beta} r)$,
(\ref{feq1})-(\ref{feq3}) become one coupled second-order
equation
\begin{eqnarray}\label{Q-eq}
\frac{d^2Q}{dr^{*2}}
+\Big[\omega^2-V_{\rm Q}(r)\Big]Q
&=&\frac{6i\omega Mf(r)}{r^5}\psi,
\end{eqnarray}
where $ V_{\rm Q}(r)$ represents the Regge-Wheeler potential~\cite{Regge:1957td}  for odd-parity perturbation as
\begin{equation}
V_{\rm Q}(r)=f(r)\Big(\frac{\lambda}{r^2}-\frac{6M}{r^3}\Big).
\end{equation}
Also, the linearized equation
(\ref{pertphi-2}) for the scalar $\delta \phi_2$ becomes a
coupled second-order equation
\begin{eqnarray}\label{psi-eq}
\frac{d^2\psi}{dr^{*2}}+\Big[\omega^2-V_{\rm \psi}(r)\Big]\psi
=-\frac{3 i M \tilde{\beta}^2(l+2)(l+1)l(l-1) f(r)}{\omega r^5}Q,
\end{eqnarray}
where $V_{\rm \psi}(r)$ denotes a potential for $\psi$
\begin{equation}
V_{\rm \psi}(r)=f(r)\Big\{\frac{\lambda}{r^2}
\Big(1+\frac{18M^2 \tilde{\beta}^2}{r^6}\Big)+\frac{2M}{r^3}\Big\}.
\end{equation}
Eqs.(\ref{Q-eq}) and (\ref{psi-eq}) represent an important
property of CS coupling to the scalar $\phi_2$.

For $s(l=0)$-mode, one has a decoupled equation for $\psi$  from (\ref{psi-eq}) as
\begin{equation}
\frac{d^2\psi}{dr^{*2}}+\Big[\omega^2-V_{\rm \psi}^{l=0}(r)\Big]\psi=0,
\end{equation}
where the potential takes the form
\begin{equation}
V_{\rm \psi}^{l=0}(r)=f(r) \Big[\frac{2M}{r^3}\Big].
\end{equation}
The $s(l=0)$-mode of $\psi$ is stable because $V_{\rm \psi}^{l=0}(r)$ is positive definite outside the horizon.
The $l=1$-mode equation leads to
\begin{equation}
\frac{d^2\psi}{dr^{*2}}+\Big[\omega^2-V_{\rm \psi}^{l=1}(r)\Big]\psi=0,
\end{equation}
where the corresponding  potential is given by
\begin{equation}
V_{\rm \psi}^{l=1}(r)=f(r)\Big[\frac{2}{r^2}\Big(1+\frac{18M^2 \tilde{\beta}^2}{r^6}\Big)+ \frac{2M}{r^3}\Big].
\end{equation}
Also, the $l=1$-mode of $\psi$ is stable because $V_{\rm \psi}^{l=1}$ is positive definite outside the horizon.
Actually, the $l=1$-mode is important because if one includes
$V(\phi_2)=\mu^2\phi_2^2/2$, it corresponds to the most prominent superradiant instability in the rotating black hole~\cite{Macedo:2018txb}.
Furthermore, it proved that all higher modes with $l\ge2$ are stable by solving  two coupled equations (\ref{Q-eq}) and (\ref{psi-eq}).
The {\it stability} of Schwarzschild black hole in the Chern-Simons coupling is closely related to   the {\it disappearance } of  black holes with scalar hair $\phi_2(r)$.
We will explore this connection in the next section.
\section{Scalarized Black holes}
Let us  develop scalarized black hole solutions by making use of  (\ref{eq-1})-(\ref{eq-3}).
Considering a spherically symmetric line element
\begin{equation}
ds^2_{\rm sBH}=-A(r)dt^2+\frac{dr^2}{B(r)}+r^2(d\varphi_1^2+\sin^2\varphi_1^2 d\varphi_2^2).
\end{equation}
From (\ref{eq-1}), we display its $(t, t)$ and $(r, r)$ components
\begin{eqnarray}\label{gral-eq1}
2\Big[r&+&4\alpha(1-3B)\phi_1\phi_1'\Big]+\frac{1}{2}\Big[B\left(4+r^2(\phi_1'^2+\phi_2'^2)\right)-4\Big]\nonumber\\
&-&16\alpha B(B-1)\left(\phi_1'^2+\phi_1\phi_1''\right)=0,\label{gral-eq2}\\
2A'\Big[r&+&4\alpha(1-3B)\phi_1\phi_1'\Big]-\frac{A}{2B}\Big[4+B(-4+r^2(\phi_1'^2+\phi_2'^2))\Big]=0.
\end{eqnarray}
Two scalar equations  (\ref{eq-2}) and (\ref{eq-3}) are given by
\begin{eqnarray}\label{scal-eq1}
\phi_1''&+&\left(\frac{2}{r}+\frac{A'}{2A}+\frac{B'}{2B}\right)\phi_1'-\frac{4\alpha\phi_1A'}{r^2A^2B}\left(B^2A'+A B'-A'B-3ABB'\right)\nonumber\\
&+&\frac{8\alpha\phi_1}{r^2A}\left(B-1\right)A''=0,\\
\phi_2''&+&\left(\frac{2}{r}+\frac{A'}{2A}+\frac{B'}{2B}\right)\phi_2'=0.\label{scal-eq2}
\end{eqnarray}
In the above,  we  observe the absence of Chern-Simons terms (${}^*RR=C_{\mu\nu}=0$), implying  that there is no source to develop scalar hair $\phi_2(r)$.
We note that (\ref{gral-eq1})-(\ref{scal-eq2}) with $\phi_2=0$ reduce  to the corresponding
equations in Ref.~\cite{Minamitsuji:2018xde}, where  black holes with scalar hair $\phi_1(r)$ was obtained  in the EGBS  theory.

In this section, we are interested in obtaining  scalarized black holes with two scalar hairs $\phi_1$ and $\phi_2$.
Near the horizon, one may consider power-series expansion of solution in terms of ($r-r_h$) as
\begin{eqnarray}\label{expan-eq}
A(r)&=&\sum^\infty_{n=1} a_n(r-r_h)^n,~B(r)=\sum^\infty_{n=1}b_n(r-r_h)^n, \nn\\
\phi_1(r)&=&\sum^\infty_{n=0}\phi_{1,n}(r-r_h)^n,\quad \phi_2(r)=\sum^\infty_{n=0}\phi_{2,n}(r-r_h)^n. \label{nhex-eq}
\end{eqnarray}
where $\{a_n, b_n,\phi_{1,n},\phi_{2,n}\}$ are constant coefficients. Substituting (\ref{expan-eq}) into (\ref{gral-eq1})-(\ref{scal-eq2}),
$b_1$, $\phi_{1,1}$ and  $a_2$  can be solved for $r_h$, $\phi_{1,0}$ and $\alpha$ as
\begin{eqnarray}
&&b_1=\frac{r_h\left(r_h^2-\sqrt{r_h^4-384\alpha^2\phi_{1,0}^2}\right)}{192\phi_{1,0}^2\alpha^2} ,\quad \phi_{1,1}=-\frac{r_h^2-\sqrt{r_h^4-384\alpha^2\phi_{1,0}^2}}{8\alpha \phi_{1,0}},\nonumber\\
&& \phi_{2,1}=\phi_{2,2}=\cdots=0,\quad  \phi_{2,0}={\rm const}
\end{eqnarray}
and so on. $a_2$ does not display here because of its complicated form.  Hence, we check that  the scalar field $\phi_2$ becomes trivial because it is const. This means that it is hard to construct scalarized black holes with scalar hair $\phi_2(r)$.
Actually,  the $n=0$ scalarized black hole  is found in Fig. 1.
Here, we choose  a rescaling of
$A(r)$ such  that $a_1$ approaches $1/2$ rather than 1 for clarity.
We note that even if one introduces a mass term of $V(\phi_2)=\mu^2 \phi_2^2/2$ in the action (\ref{Action}), any scalarized black holes with $\phi_2(r)$ are not allowed because there is no way of escaping from the no-hair theorem~\cite{Garcia:2018sjh}.

\begin{figure}[htb]
\centering
\includegraphics{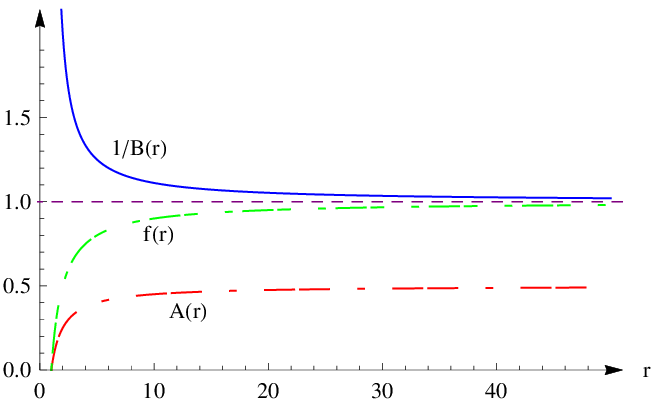}
\hfill%
\includegraphics{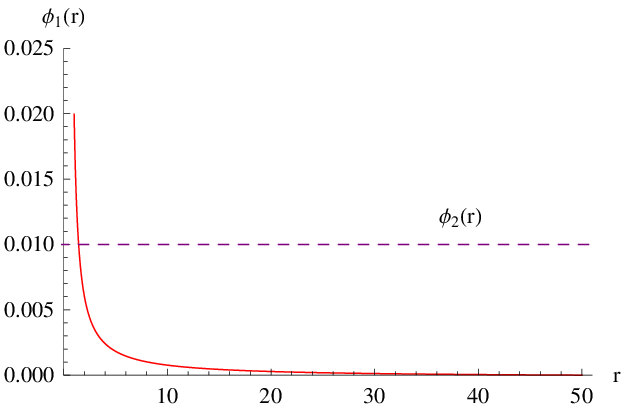}
\caption{Metric functions [$ A(r), f(r),1/B(r)$] and scalar fields [$\phi_1(r), \phi_2(r)$] as functions of  $r\in [1,50]$ for the $n=0$ scalarized  black hole.
The horizon is located at $r=r_h=1$ and $\alpha$ is chosen $0.092>\alpha_{\rm th}=0.0907$ for the $n=0$ scalarized black hole.}\label{fig1}
\end{figure}
Finally, let  us discuss the stability issue of the $n=0$ scalarized black hole in the GBCSS theory.
As was mention in~\cite{Blazquez-Salcedo:2018jnn}, the difference between exponential and quadratic couplings in the EGBS theory is that the $n=0$  scalarized black hole is stable for the exponential coupling, while the $n=0$ scalarized black hole is unstable for the quadratic coupling.
Recently,  it is shown that the quadratic term controls the onset of the instability giving the $n=0$ scalarized black hole, while the higher-order  coupling terms  including the exponential coupling control the nonlinearities quenching the instability and thus, control the stability of the $n=0$ black hole~\cite{Silva:2018qhn}.
Therefore, we expect that the $n=0$ scalarized black hole is unstable against perturbations in the GBCSS theory since this theory has a quadratic coupling term.

\section{Discussions}

We state that the appearance of scalarized black holes is directly related to the instability of black hole without scalar hair.
We have found that the Schwarzschild (static) black hole is unstable against the $s(l=0)$-mode perturbation of $\delta \phi_1$ coupled to the Gauss-Bonnet term, implying  the appearance of scalarized black holes in the EGBS theory. On the other hand, the static black hole is stable against  all modes  of $\delta \phi_2$ coupled to the Chern-Simons term, implying  the disappearance of scalarized black holes in the ECSS theory.  This is so because the Chern-Simons term
could not activate a scalar monopole field $\phi_2(r)$ in the static black hole spacetime. This explains why the scalarized static black holes could not be found in the ECSS theory.

In the single scalar coupling of $\phi_1=\phi_2=\phi$, choosing $\bar{\phi}=0$ provides the Schwarzschild solution.
In this case, the linearized equations around the black hole lead to those of the EGBS theory, suggesting the appearance of scalarized black holes.
In this case, the role of the Chern-Simons  term disappears.

Furthermore, we note that the  Kerr black hole with scalar hair could be obtained from the superradiant instability of Kerr black hole in the Einstein-Klein-Gordon theory~\cite{Herdeiro:2014goa}.
Here the  threshold instability is given by the $n=0,~l=m$ modes of a perturbed  scalar and these $l=m$ clouds can be promoted to Kerr with scalar hair  in the full Einstein-Klein-Gordon theory.
Also, the non-perturbative spinning black holes  could be obtained from the ECSS (linear coupling) theory~\cite{Delsate:2018ome}.
Its linearized scalar equation takes the form of $\bar{\nabla}^2\delta\phi_2+ \beta \delta({}^{*}RR) =0$ in the Kerr black hole background.
In this case, however, we do not know the stability issue of Kerr black hole and thus, connection between non-perturbative spinning black holes and instability of Kerr black hole is missed.

In our model (\ref{Action}),  we  may obtain  the Kerr black hole solution to (\ref{eq-1})-(\ref{eq-3}) when setting $\bar{\phi}_1=\bar{\phi}_2=0$  with $a=J/M$ because
\begin{equation}
{}^{*}\bar{R}\bar{R}=\frac{96aM^2r\cos \varphi_1(3r^2-a^2\cos^2\varphi_1)(r^2-3a^2\cos^2\varphi_1)}{(r^2+a^2\cos^2\varphi_1)^6}\not=0
\end{equation}
in Boyer-Lindquist coordinates.
In this case, the linearized Einstein equation ($\delta R_{\mu\nu}=0$) is completely decoupled from the perturbed scalar $\delta \phi_2$, indicating no instability from metric perturbations. A relevant equation is  the linearized scalar equation  of $(\bar{\nabla}^2+2 \beta{}^{*}\bar{R}\bar{R} )\delta\phi_2=0$, which is a Teukolsky-like equation with an effective mass of $-2 \beta{}^{*}\bar{R}\bar{R}$.
For $\beta \le 1$, recently, the Kerr black hole may be shown to be unstable against the $l=m=1$ and $l=m=2$  modes of $\delta \phi_2$ for the spin parameter $a=0.9$~\cite{Gao:2018acg}.
The authors in~\cite{Gao:2018acg} insisted that this modal instability  is nothing to do with the superradiant instability and thus,  scalarized Kerr black holes would be found from this model. These would-be solutions should be compared to scalarized Kerr black holes obtained in Einstein-Klein-Gordon theory~\cite{Herdeiro:2014goa}.
Here, the spontaneous scalarization  may be possible because the Chern-Simons term gives different results only in the rotating solution of parity-odd source.
However, one unclear issue is  why the $s$-mode of $l=m=0$ was not chosen for the modal stability analysis in this approach.
 \vspace{1cm}

{\bf Acknowledgments}
\vspace{1cm}

This work was supported by the National Research Foundation of Korea (NRF) grant funded by the Korea government (MOE)
 (No. NRF-2017R1A2B4002057).

 \vspace{1cm}

\newpage
\section*{Appendix: Ten perturbation equations}
\begin{eqnarray}
(t,t):&& e^{-i\omega t}E_1Y=0\nn\\
(t,r):&& e^{-i\omega t}E_2Y=0\nn\\
(t,\varphi_1):&&e^{-i\omega t}\Big(E_3
\partial_{\varphi_1}Y
+O_1\partial_{\varphi_2}Y\Big)=0\nn\\
(t,\varphi_2):&&e^{-i\omega t}\Big(E_3 \partial_{\varphi_2}Y
+O_2\partial_{\varphi_1}Y\Big)=0\nn\\
(r,r);&&e^{-i\omega t}E_4 Y=0\nn\\
(r,\varphi_1):&&e^{-i \omega t}\Big(E_5
\partial_{\varphi_1}Y
+O_3\partial_{\varphi_2}Y\Big)=0\label{comps} \\
(r,\varphi_2):&&e^{-i\omega t}\Big(E_5
\partial_{\varphi_2}Y
+O_4\partial_{\varphi_1}Y\Big)=0\nn\\
(\varphi_1,\varphi_1):&&e^{-i\omega t}\Big(E_{6}Y
+E_{7}\partial_{\varphi_1}^2Y+O_{5}\partial_{\varphi_2} Y
+O_{6}\partial_{\varphi_1}\partial_{\varphi_2}Y\Big)=0\nn\\
(\varphi_1,\varphi_2):&&e^{-i\omega t}\Big(E_{8}\partial_{\varphi_2}Y
+E_{7}\partial_{\varphi_1}\partial_{\varphi_2}Y
+O_7Y+O_8\partial_{\varphi_1}^2Y\Big)=0\nn\\
(\varphi_2,\varphi_2):&&e^{-i \omega t}\Big(E_{9}Y
+E_{7}\partial_{\varphi_2}^2Y+E_{10}\partial_{\varphi_1}Y
+O_{9}\partial_{\varphi_2}Y
+O_{10}\partial_{\varphi_1}\partial_{\varphi_2}Y\Big)=0 \nn.
\end{eqnarray}
The perturbation equations
(\ref{comps}) imply twenty constraints like
\begin{eqnarray}
E_{i}=0,~~~O_{i}=0,~~~ {\rm for}~i=1,\cdots,10.
\end{eqnarray}
The explicit forms of $E_i$ and $O_i$ are as follows:
\begin{eqnarray}
&&
E_1=\frac{1}{2r^5}\Big[2r^5\omega^2K(r)+r(-2M^2f^{-1}+r^2\lambda)H_0(r)
+2i(3M-2r)r^3\omega H_1(r)\nn\\
&&\hspace*{3em}-rf(2M^2-r^4\omega^2)H_2(r)+2Mr^3fK^{\prime}(r)
+r^3(5M-2r)H_0^{\prime}(r)\nn
\nn\\&&\hspace*{3em}-2ir^5f\omega H_1^{\prime}(r)-Mr^3f^2H_2^{\prime}(r)
-r^5fH_0^{\prime\prime}(r)\Big],\nn\\
&& E_2=\frac{i}{2r^2}\Big[2(r-3M)\omega f^{-1}
K(r)-2\omega rfH_2(r)+4\omega r^2K^{\prime}(r)-i\lambda H_1(r)\Big],\nn\\
&& E_3=\frac{i}{2r^2}\Big[\omega r^2K(r)-2iMH_1(r)+\omega r^2fH_2(r)-ir^2
fH_1^{\prime}\Big],\nn\\
&& E_4=\frac{1}{2r^4f^2}\Big[2M(2r-3M)f^{-1}H_0(r)+2iM\omega r^2H_1(r)
-f\Big(6M^2-4Mr\nn\\
&&\hspace*{3em}+\omega^2r^4-r^2f\lambda\Big)H_2(r)-2r(6M^2-7Mr+2r^2)K^{\prime}(r)
-Mr^2H_0^{\prime}
\nn\\
&&\hspace*{3em}+2i\omega r^4fH_1^{\prime}+r
f(6M^2-7Mr+2r^2)H_2^{\prime}-2r^4f^2K^{\prime\prime}(r)
+r^4fH_0^{\prime\prime}\Big],\nn\\
 &&
E_5=\frac{-1}{2r^3f}\Big[(r-M)rf^{-1}H_0(r)-i\omega r^3H_1(r)
-(2M^2-3Mr+r^2)H_2(r)\nn\\
&&\hspace*{3em}+r^3fK^{\prime}(r)-r^3H_0^{\prime}\Big],\nn\\
&& E_6=\frac{-1}{2r^2}\Big[2Mrf^{-1}H_0(r)-2i\omega r^3H_1(r)+2
(2M^2+Mr-r^2)H_2(r)\nn\\
&&\hspace*{3em}+r^2(\omega^2r^2f^{-1}+\lambda+2)K(r)
-2r^2(3M-2r)K^{\prime}(r)-r^3H_0^{\prime}(r)
\nn\\
&&\hspace*{3em}-r^3f^2H_2^{\prime}(r)+r^4fK^{\prime\prime}(r)\Big],\nn\\
&& E_7=
\frac{1}{2r}\Big[rf^{-1}H_0(r)-rfH_2(r)\Big],~E_8=-E_7\cot\varphi_1,
~E_{9}=E_{6}\sin^2\varphi_1, \nn \\
&& E_{10}=E_{7}\cos\varphi_1\sin\varphi_1,\nn
\end{eqnarray}
\begin{eqnarray}
&&\hspace*{-1.5em}
O_1=\frac{\csc\varphi_1}{2r^3}\Big[ir^2f\Big\{2\omega h_1(r)+\omega rh_1^{\prime}(r)
-irh_0^{\prime\prime}(r)\Big\}+(4M-\lambda r)h_0(r)\nn\\
&&\hspace*{5em}-\frac{3\tilde{\beta}}{r^2}Mf\Big\{2\psi(r)-r\psi^{\prime}(r)\Big\}\Big],\nn\\
&&\hspace*{-1.5em} O_2=\frac{-1}{2r^3}\Big[ir^2f\sin\varphi_1
\Big\{2\omega h_1(r)+\omega rh_1^{\prime}(r)-irh_0^{\prime\prime}(r)\Big\}
-(rf+2M\cos2\varphi_1)\csc\varphi_1h_0(r)\nn\\
&&\hspace*{4em}+r(\csc\varphi_1-\lambda\sin\varphi_1)h_0(r)
-\frac{3\tilde{\beta}}{r^2}Mf\sin\varphi_1\Big\{2\psi(r)-r\psi^{\prime}(r)\Big\}\Big],\nn\\
&&\hspace*{-1.5em}
O_3=\frac{\csc\varphi_1}{2r^3f}\Big[2i\omega r^2h_0(r)-i\omega r^3h_0^{\prime}(r)+
(2rf+\omega^2r^3-\lambda rf)h_1(r)-\frac{3i\tilde{\beta}}{r}kM\psi(r)\Big],\nn\\
&&\hspace*{-1.5em}
O_4=\frac{-1}{2r^3f}\Big[i\omega r^2\sin\varphi_1\{2h_0(r)-rh_0^{\prime}\}
+\Big\{-rf\cos\varphi_1\cot\varphi_1+(rf+\omega^2r^3)\sin\varphi_1\nn\\
&&\hspace*{5em}+rf(\csc\varphi_1-\lambda\sin\varphi_1)\Big\}h_1(r)-\frac{3i\tilde{\beta}}{r}kM\sin\varphi_1\psi(r)\Big],\nn\\
&&\hspace*{-1.5em}
O_5=\frac{\csc\varphi_1\cot\varphi_1}{r^3f}
\Big[i\omega r^3h_0(r)+rf\Big\{2Mh_1(r)+r^2fh_1^{\prime}(r)\Big\}\Big],\nn\\
&&\hspace*{-1.5em}
O_6=-O_5\tan\varphi_1,\nn\\
&&\hspace*{-1.5em}
O_7=\frac{ O_5 \lambda}{2}\Big[\sin^2\varphi_1\tan\varphi_1\Big],\nn\\
&&\hspace*{-1.5em}
O_8=O_5\sin^2\varphi_1\tan\varphi_1,\nn\\
&&\hspace*{-1.5em}
O_9=-O_5\sin^2\varphi_1,\nn\\
&&\hspace*{-1.5em} O_{10}=O_8\nn
\end{eqnarray}
with $\tilde{\beta}=8\bar{\phi}_2 \beta$.

\end{document}